\documentclass[10pt]{article}

\usepackage{amstext,amsgen,latexsym,amsmath}
\usepackage{amstext,amssymb,amsfonts,latexsym}
\usepackage{theorem}
\usepackage{pifont}
\usepackage{graphics,epsfig}

 \newcommand{\bs}{\bigskip}
 \newcommand{\ms}{\medskip}
 \newcommand{\n}{\noindent}
 \newcommand{\s}{\smallskip}
 \newcommand{\hs}[1]{\hspace*{ #1 mm}}
 \newcommand{\vs}[1]{\vspace*{ #1 mm}}

 \newcommand{\setempty}{\mathrm{\O}}

 \newcommand{\nat}{\mathbb{N}}
 
 \newcommand{\integer}{\mathbb{Z}}

 \newcommand{\ie}{\textrm{i.e.},\hspace*{2mm}}
 \newcommand{\eg}{\textrm{e.g.},\hspace*{2mm}}
 
 \newcommand{\etalc}{\textrm{et al.}}

 \newcommand{\AAA}{{\cal A}}
 \newcommand{\BB}{{\cal B}}
 
 \newcommand{\CC}{{\cal C}}

 \newcommand{\MM}{{\cal M}}

 \newcommand{\p}{\mathrm{P}}

 \newcommand{\bpp}{\mathrm{BPP}}

 \newcommand{\bqp}{\mathrm{BQP}}

 \newcommand{\poly}{\mathrm{poly}}

 \def\bbox{\vrule height6pt width6pt depth1pt}

\theoremstyle{plain}
\theoremheaderfont{\bfseries}
 \newtheorem{theorem}{Theorem}[section]
 \newtheorem{lemma}[theorem]{Lemma}
 \newtheorem{proposition}[theorem]{Proposition}

 {\theorembodyfont{\rmfamily}
}
 {\theorembodyfont{\rmfamily} }
 {\theorembodyfont{\rmfamily} }

 \newenvironment{proof}{\par \noindent
            {\bf Proof. \hs{2}}}{\hfill$\Box$ \vspace*{3mm}}

 \newenvironment{proofof}[1]{\vspace*{5mm} \par \noindent
         {\bf Proof of #1.\hs{2}}}{\hfill$\Box$ \vspace*{3mm}}

 \newcommand{\ceilings}[1]{\lceil #1 \rceil}
 \newcommand{\floors}[1]{\lfloor #1 \rfloor}
 \newcommand{\pair}[1]{\langle #1 \rangle}

 \newcommand{\qubit}[1]{| #1 \rangle}

\newif\ifnotesw\noteswtrue
   {\ifnotesw\marginpar[\hfill\(\top\)]{\(\top\)}\fi}%
      {\ifnotesw\marginpar[\hfill\(\bot\)]{\(\bot\)}\fi}
      
\newcommand{\mnote}[1]%
   {\ifnotesw\marginpar%
          [{\scriptsize\begin{minipage}[t]{\marginparwidth}
          \raggedleft#1%
                  \end{minipage}}]%
          {\scriptsize\begin{minipage}[t]{\marginparwidth}
          \raggedright#1%
                  \end{minipage}}%
    \fi}

\newcommand{\ignore}[1]{}

\begin{document}
\pagestyle{plain}
\begin{center}
{\Large {\bf An Algorithmic Argument for \s\\
Nonadaptive Query Complexity Lower Bounds on \ms\\
Advised Quantum Computation}}
\footnote{This work was in part supported by 
the Natural Sciences and Engineering Research Council of Canada. An extended abstract 
will appear in the Proceedings of the 29th International Symposium on Mathematical 
Foundations of Computer Science, Lecture Notes in Computer Science, Springer-Verlag, 
Prague, August 22--27, 2004.} \bs\bs\\

\begin{tabular}{c@{\hspace{20mm}}c}
{\sc Harumichi Nishimura} & {\sc Tomoyuki Yamakami}
\end{tabular}\ms\\

{Computer Science Program, Trent University} \\
{Peterborough, Ontario, Canada  K9J 7B8} \ms\\
\end{center}
\bs 

\n{\bf Abstract.}\hs{1} 
This paper employs a powerful argument, called an algorithmic 
argument, to prove lower bounds of the quantum query complexity 
of a multiple-block ordered search problem in which, 
given a block number $i$, we are to find a location of a 
target keyword in an ordered list of the $i$th block. Apart from much 
studied polynomial and adversary methods for quantum query 
complexity lower bounds, our argument shows that the multiple-block 
ordered search needs a large number of nonadaptive oracle queries 
on a black-box model of quantum computation that is also 
supplemented with advice. Our argument is also applied to the 
notions of computational complexity theory: quantum truth-table 
reducibility and quantum truth-table autoreducibility.  

\ms
\n{\sf Keywords:} algorithmic argument, query complexity, 
nonadaptive query, advice, quantum computation, ordered search

\section{An Algorithmic Argument for Query Complexity}
\label{sec:argument}

A major contribution of this paper is the demonstration of a 
powerful argument, which we refer to {\em algorithmic argument}, 
to prove a lower bound of the nonadaptive query complexity for 
a multiple-block ordered search problem on advised quantum 
computation. In the literature, quantum query complexity lower 
bounds have been proven by classical adversary methods \cite{BBBV97}, 
polynomial methods \cite{BBCMW01}, and quantum adversary methods 
\cite{Amb02,BSS03,HNS02}. Each method has its own strength and 
advantages over its simplicity, clarity, and dexterity. An 
algorithmic argument, however, is essentially different from 
these methods in its constructive manner. A basic scheme of our 
algorithmic argument is illustrated as follows: we (i) commence 
with the faulty assumption that a quantum algorithm $\AAA$ of 
low query complexity exists, (ii) define a compression scheme 
$E$ that encodes each input $s$ of fixed length into a shorter 
string $E(s)$, and (iii) prove the one-to-oneness of $E$ by 
constructing a decoding algorithm from $\AAA$ that uniquely 
extracts $s$ from $E(s)$, which leads to a conspicuous contradiction 
against the pigeonhole principle\footnote{The pigeonhole principle 
formalizes the intuition that, when $n$ pigeons rest in fewer than $n$ nests, 
at least two pigeons share the same nest.}. In this paper, we build a classical 
algorithm that decodes $E(s)$ by simulating $\AAA$ in a deterministic fashion. We therefore reach the conclusion that 
any quantum algorithm should require high query complexity. When $\AAA$ further 
models ``uniform'' quantum computation, we can use resource-bounded Kolmogorov 
complexity\footnote{The use of Kolmogorov complexity unintentionally adds an 
extra constant additive term and thus gives a slightly weaker lower bound.} 
as a technical tool in place of the pigeonhole principle. 

We apply our algorithmic argument to obtain a new nonadaptive 
query complexity lower bound on a query computation model, 
known as a {\em black-box quantum computer} (sometimes called a 
{\em quantum network} \cite{BBCMW01}), in which a {\em query} 
is an essential method to access information stored outside of 
the computer. The minimal number of such queries, known as the 
{\em query complexity}, measures the smallest amount of information 
necessary to finish the desired computation. Query complexity 
lower bounds on 
various quantum computational models have been studied for numerous 
problems, including ordered search \cite{Amb99,BW99,FGGS98,HNS02}, 
unordered search \cite{Amb02,BBCMW01,BBHT98,BBBV97}, element 
distinctness \cite{Amb03v2,BDH+01,Shi02}, and collision \cite{Aar02,Amb03v2,Shi02}. 

A black-box quantum computer starts with a fixed initial state 
(\eg $\qubit{0\cdots 0}$), accesses a given source $x$ (which is 
called an ``oracle'') by way of {\em queries}---``what is the binary 
value at location $i$ in $x$?''---and outputs a desired solution 
with small error probability. If any query (except the first one) 
is chosen according to the answers to its previous queries, such 
a query pattern is conventionally referred to as {\em adaptive}. 
Adaptive oracle quantum computation has been extensively studied 
and have given rise to useful quantum algorithms, e.g., 
\cite{BV97,CCDFGS03,DJ92,Gro96,Sim97}. 
An adaptive computation in general requires a large number of times of 
interactions between the computer and a given oracle.  Since a quantum 
computer is known to be sensitive to any interaction with other physical 
systems, such as an oracle, it would be desirable to limit the 
number of times of interactions with the oracle. In contrast, the query 
pattern in which all the query words are prepared before the 
first query is referred to as {\em nonadaptive queries} (including 
{\em parallel queries} and {\em truth-table queries}). Recently, 
Buhrman and van Dam \cite{BD99} and Yamakami \cite{Yam03} extensively 
studied the nature of parallel queries on quantum computation. 
By visiting the results in \cite{BB94,BD99,CEMM97,DJ92,Sim97,Yam03}, 
we can find that quantum nonadaptive queries are still more powerful 
than classical adaptive queries. This paper pays its attention to the truth-table 
query model in which the query words produced in a prequery quantum state 
are answered all at once. 

Our black-box quantum computation is further equipped with 
{\em advice}, which was first discussed by Karp and Lipton \cite{KL82}, to 
provide an additional source of information that boosts the computational 
power. Such advice supplements the information drawn from an oracle and 
therefore the advice reduces the query complexity. 
Lately, time-bounded advised quantum computation was 
introduced by Nishimura and Yamakami \cite{NY04} and discussed 
later by Aaronson \cite{Aar04}. Their notion has a close connection 
to nonuniform computation \cite{NY04} and also one-way communication 
\cite{Aar04}. Our interest in this paper lies in the relationship between the 
size of advice and the nonadaptive quantum query complexity. 

Based on a black-box model of quantum computation with advice, we 
employ our algorithmic argument to prove a lower bound of the query 
complexity of a so-called {\em ordered search problem}. For 
simplicity, we focus our interest on the ordered search problem 
of the following kind: given an $N$-bit string $x$ of the form 
$0^{j-1}1^{N-j+1}$ for certain positive integers $N$ and $j$, we are 
to find the leftmost location $s$ of $1$ (which equals $j$). 
Such a unique location $s$ is called the {\em step} of $x$ (since 
the input $x$ can be viewed as a so-called {\em step function}). 
This ordered search problem is one of the well-studied problems for 
their quantum adaptive query complexity.  Naturally, we can expand 
this problem into a ``multiple-block'' ordered search problem, in 
which we are to find the step (called the $i$th step) in each block $i$  
when the block number 
$i$ is given as an input. To avoid the reader's confusion, we call the 
standard ordered search problem the {\em single-block ordered search 
problem}. This paper presents a new query complexity lower 
bound for the multiple-block ordered search problem on a nonadaptive 
black-box quantum computer with the help of advice.

Independently, Laplante and Magniez \cite{LM04} also found a similar 
algorithmic argument to demonstrate general lower bounds of 
randomized and quantum query complexity. Their argument, nonetheless, 
is meant for the adaptive query complexity without advice strings and is 
different in its nature of query computation from our argument. 

Finally, we note that an algorithmic argument is not new in 
classical complexity theory. Earlier, Feigenbaum, Fortnow, Laplante, 
and Naik \cite{FFLN98} applied an algorithmic argument to show that 
the multiple-block ordered search problem is hard to solve on a 
classical Turing machine using nonadaptive queries. Their proof, 
nonetheless, cannot be directly applied to the case of black-box 
quantum computation since their proof exploits the fact that a 
probabilistic polynomial-time Turing machine with polynomial advice 
can be simulated by a certain deterministic polynomial-time Turing 
machine with polynomial advice. Our technique developed in this paper, 
to the contrary, enables us to show a desired quantum query complexity 
lower bound for the multiple-block ordered search problem.  

In the subsequent sections, we will give the formal definition of 
nonadaptive black-box query computation and of the 
multiple-block ordered search 
problem. We will also present an overview of our lower bound results 
before proving the main theorems. Finally, we hope that an algorithmic 
argument would find more useful applications in other fields of theoretical 
computer science in the future.

\section{A Model of Nonadaptive Query Computation}

We formally describe a black-box model of quantum nonadaptive query 
computation. The reader may refer to \cite{BBCMW01} for the formal 
description of quantum ``adaptive'' query computation. In particular, 
we use a ``truth-table'' query model rather than the ``parallel'' 
query model given in \cite{BD99,Yam03} to simplify our algorithmic 
argument although our results still hold in the parallel query model. Note that 
these two models are fundamentally equivalent in 
the classical setting since there is no {\em timing problem} as it occurs in 
the quantum case (see, \eg \cite{Yam03} for more details). In our 
truth-table query model, all queries are made at once after the first 
phase of computation and the second phase leads to a desirable solution 
without any query. This truth-table query model can be seen as a 
special case of the parallel query model of Yamakami \cite{Yam03}. 

In the rest of this paper, we assume the reader's familiarity with 
the fundamental concepts in computational complexity theory  
and quantum computing (see, \eg \cite{DK00} for computational 
complexity theory and \cite{NC00} for quantum computing). 
Hereafter, we fix our alphabet $\Sigma$ to be 
$\{0,1\}$ for simplicity. Let $\nat$ be the set of all {\em natural 
numbers} (\ie nonnegative integers) and set 
$\nat^{+}=\nat- \{0\}$. For any two integers $m$ and $n$ with 
$m<n$, the notation $[m,n]_{\integer}$ denotes the set 
$\{m,m+1,m+2,\ldots,n\}$. For any 
$n\in\nat^{+}$ and $i\in[1,2^n]_{\integer}$, let 
$\mathrm{bin}_n(i)$ represent the lexicographically $i$th binary 
string in $\Sigma^n$ (\eg $\mathrm{bin}_n(1)=0^n$ and 
$\mathrm{bin}_n(2^n)=1^n$). For any finite set $A$, write $|A|$ for 
the cardinality of $A$. The {\em tower of 2s} is 
inductively defined by $2_0=1$ 
and $2_n=2^{2_{n-1}}$ for each $n\in\nat^+$. For convenience, let $\mathrm{Tower2}=\{2_n\mid n\in\nat\}$. 
All {\em logarithms} are to base two and all {\em polynomials} have integer coefficients. 
For convenience, we set $\log{0}=0$.

Fix $N$ as a power of 2, say, $N=2^n$ for a certain number $n\in\nat^{+}$. 
A {\em (black-box) problem}\footnote{To simplify later arguments, 
we consider only the case where $N$ is a power of 2 although it is 
possible to discuss the general case of $N$'s taking any other integer.} 
$F_N$ over alphabet $\Sigma$ is a function that maps each instance $f$ to 
its value $F_N(f)$, where $f$ is any function from $\Sigma^n$ to 
$\Sigma$. For convenience, $f$ is often identified with its 
characteristic sequence\footnote{A {\em characteristic sequence} 
of $f$ is $x_1x_2\cdots x_N$, where the $i$th bit $x_i$ is the 
value $f(\mathrm{bin}_n(i))$ for each $i\in[1,N]_{\integer}$.} 
of length $N$.  

To solve such a problem $F_N$, we carry out a nonadaptive query 
quantum computation on a black-box quantum computer. A 
{\em nonadaptive black-box quantum computer} is formally described as a 
series of pairs of unitary operators, say $\{(U_n,V_n)\}_{n\in\nat}$. 
Let $n$ be any fixed size of our instances. The quantum computer 
$(U_n,V_n)$ consists of three registers. The first register is 
used to generate query words, the second register is used to 
receive the oracle answers to these query words, and the third 
register is used to perform non-query computation. Assume that an 
instance $x\in\Sigma^{N}$ of the problem $F_N$ is given to the 
computer as an oracle and all the registers are initially set to 
be the quantum state $\qubit{0}\qubit{0}\qubit{0}$.  In the first phase, the computer 
changes the initial quantum state $\qubit{0}\qubit{0}\qubit{0}$ to the {\em prequery 
quantum state} $\sum_{i_1,\ldots,i_T} \qubit{i_1,\ldots,i_T}
\qubit{0^T}\qubit{\phi_{i_1,\ldots,i_T}}$ 
by applying the unitary operator $U_n$, where $T\in\nat$ and 
$i_1,\ldots,i_T\in[1,N]_{\integer}$. Each index $i_j$ is called a 
{\em query word} and the list\footnote{We assume that any query list 
consists of exactly $T$ query words along every computation path of 
nonzero amplitude; however, any query list may contain duplicated query 
words. Such an assumption simplifies the proofs of our main theorems.}  
$\vec{i}=(i_1,i_2,\ldots,i_T)$ is called a {\em query list}. Strictly speaking, 
each query word $i_j$ in a query list $\vec{i}$ should be generated as 
the $n$-bit string $\mathrm{bin}_n(i)$ in the first register. In this way, 
we often identify $[1,N]_{\integer}$ with $\Sigma^{\log{N}}$. We next 
prepare the unitary 
operator $O_x$ that represents $x$. The application of $O_x$ then 
results in the {\em postquery quantum state} 
$\sum_{i_1,\ldots,i_T}\qubit{i_1,\ldots,i_T}
\qubit{x_{i_1}\cdots x_{i_T}}\qubit{\phi_{i_1,\ldots,i_T}}$. 
In the final phase, the computer 
applies the unitary operator $V_n$ and then halts. When the 
computer halts, its output state becomes 
$V_nO_xU_n(\qubit{0}\qubit{0}\qubit{0})$.  The first $|F_N(x)|$ 
qubits of the third register are measured on the computational 
basis to obtain the outcome of the computation.

The {\em $\epsilon$-error bounded quantum nonadaptive query 
complexity} of the problem $F_N$, denoted by $Q_{\epsilon}^{tt}(F_N)$, 
is defined to be the minimal number $T$ of the nonadaptive queries 
made by any nonadaptive black-box quantum computer with oracle $x$ 
such that $F_N(x)$ is observed with error probability at most 
$\epsilon$ by the measurement of the output state. Throughout this 
paper, we restrict all the amplitudes of a bounded-error black-box 
quantum computer to the amplitude set $\{0,\pm 3/5,\pm 4/5,\pm 1\}$. 
This restriction does not affect our results since bounded-error 
quantum algorithms are known to be robust against the choice of 
an amplitude set \cite{ADH97}. 

In the case where an advice string $h_x$ is given as a supplemental input, a 
nonadaptive black-box quantum computer starts with the initial 
quantum state $\qubit{0}\qubit{0}\qubit{h_x}$ instead of 
$\qubit{0}\qubit{0}\qubit{0}$. 
We denote by $Q^{k,tt}_{\epsilon}(F_N)$ the {\em $\epsilon$-error 
bounded quantum nonadaptive query complexity of $F_N$ given advice 
of length $k$}. For convenience, we often suppress the subscript 
$\epsilon$ if $\epsilon=1/3$. 

We further expand the problem $F_N$ into a ``multiple-block'' 
problem in the following section.

\section{Multiple Block Problems}

We first give a general scheme of how to expand a (black-box) 
problem $F_N$ into a ``multiple-block'' problem $F_{M,N}$. Formally, 
for any numbers $n,M\in\nat$ and $N=2^n$, we define an {\em (black-box) 
$M$-block problem} $F_{M,N}$ as follows. Let 
$\Gamma_{M,N}=[1,M]_{\integer}\times\Sigma^{n}$, where the first 
part of $\Gamma_{M,N}$ indicates ``block numbers'' and the second 
part indicates ``locations.'' An instance to the problem $F_{M,N}$ 
is a pair $(i,f)$ of an integer $i\in[1,M]_{\integer}$ and a function $f$ 
from $\Gamma_{M,N}$ to $\Sigma$. We often abbreviate $f(i,x)$ as $f_i(x)$. 
A problem $F_{M,N}$ is a function mapping $(i,f)$ to its value 
$F_{M,N}(i,f)$. The function $f$ is given as an oracle and $i$ is 
given as an input to a black-box quantum computer. To access the 
value of $f$, we need to query a pair $(i,y)\in \Gamma_{M,N}$ (called 
a {\em query word}) to the oracle $f$. Our task is to compute 
$F_{M,N}(i,f)$. Given such a problem $F_{M,N}$, the quantum computer starts 
with the initial quantum state $\qubit{0}\qubit{0}\qubit{i}$ with a block number 
$i$, and attempts to compute the value 
$F_{M,N}(i,f)$, which depends only on $f_i$, by making queries to $f$ 
given as an oracle. (If an advice string $h$ is augmented, the initial 
quantum state must be $\qubit{0}\qubit{0}\qubit{i,h}$.) Obviously, 
the (classical/quantum and adaptive/nonadaptive) query complexity for 
any multiple-block problem $F_{M,N}$ is at most $N$. Later, we 
identify $f_i$ with its characteristic sequence of length $N$.

The {\em $M$-block ordered search problem} $G_{M,N}$, where $N=2^n$, 
is formally defined as follows. The domain of $G_{M,N}$ is the 
set $\{(i,x_1\cdots x_M)\in[1,M]_{\integer}\times\Sigma^{MN} 
\mid \forall\: j\: \exists\: s_j\: \forall\: k\:
[\ (x_j)_{k}=1\ \mbox{if }k\geq s_j\mbox{ and }
(x_j)_{k}=0\ \mbox{if }k<s_j]\}$, where each $x_j$ is taken from 
$\Sigma^N$. Each $s_j$ is called the {\em step} of $x_j$. The 
output value  $G_{M,N}(i,x_1x_2\cdots x_M)$ is the $i$th step 
$s_i$. For a later use, we define the modified problem $G_{M,N,p}$ for each 
number $p\in[1,n]_{\integer}$ as follows. The domain of $G_{M,N,p}$ is 
the same as that of $G_{M,N}$ but the outcome 
$G_{M,N,p}(i,x_1x_2\cdots x_M)$ is instead the last $p$ bits of $s_i$. 
Obviously, $G_{M,N,n}$ coincides with $G_{M,N}$. 

To solve the multiple-block ordered search problem $G_{M,N}$, our 
nonadaptive black-box quantum computer $(U_n,V_n)$ operates in the 
following fashion.  Given a pair $(i,x)$ of a number 
$i\in[1,M]_{\integer}$ and an $MN$-bit string $x=x_1x_2\cdots x_M$, 
where each $x_i$ is in $\Sigma^N$, $(U_n,V_n)$ starts with a 
block number $i$ and an advice string $h$ (which is given independent 
of $i$) of length $k$ and attempts to compute the value $G_{M,N}(i,x)$ 
with small error probability. It is desirable in practice to 
minimize the number of queries and also the length of an advice 
string.

\section{Query Complexity Lower Bounds: Overview}

This section presents an overview of the adaptive and nonadaptive 
query complexity bounds for the multiple-block ordered search 
problem $G_{M,N}$. This problem can separate the power of quantum 
adaptive query computation and that of quantum nonadaptive query 
computation. 

For its quantum adaptive query complexity, the single-block 
ordered search problem $G_{1,N}$ is one of the well-studied 
problems. In this adaptive query case, a simple binary search 
algorithm provides a trivial adaptive query complexity upper 
bound of $\log{N}$. The lower bound of the adaptive query 
complexity $Q(G_{1,N})$ was explored in \cite{BW99,FGGS98}, and 
the $\Omega(\log{N})$-lower bound was recently given by Ambainis 
\cite{Amb99} and H{\o}yer, Neerbek, and Shi \cite{HNS02}. For 
the multiple-block problem $G_{M,N}$, a trivial query complexity 
upper bound is also $\log{N}$.

\begin{center}
{\small \begin{tabular}{|l|c|c|c|c|}\hline
problem    & \multicolumn{2}{c|}{$G_{1,N}$} 
& \multicolumn{2}{c|}{$G_{M,N}$}\\ \hline
advice     & no advice & advice length $k$ 
& no advice & advice length $k$ \\ \hline 
upper bound & $N-1$ & $N/2^k-1$  & $N-1$  
& $N/2^{\floors{k/M}}-1$      \\ \hline
lower bound & $\Omega(N)$ & $\Omega(N)/2^{2k}$ 
& $\Omega(N)$   &  $\Omega(p(N,M,k))$         \\ \hline 
\end{tabular}}
\ms

Table 1: Quantum nonadaptive query complexity bounds of 
$G_{1,N}$ and $G_{M,N}$
\end{center}

On the contrary, the nonadaptive query complexity has a trivial 
upper bound of $N-1$ for the multiple-block problem $G_{1,N}$. 
Similarly, in the presence of advice of length $k$, the query 
complexity $Q^{k,tt}(G_{1,N})$ is upper-bounded by $N/2^k-1$. 
As for the lower bound, using our algorithmic argument, we can 
show in Theorem \ref{ordered-nonadaptive} a lower bound 
$Q^{k,tt}(G_{1,N})\geq \Omega(N)/2^{2k}$, which almost matches 
the above trivial upper bound for $G_{1,N}$. Turning to the 
multiple-block problem $G_{M,N}$, we can show in Theorem 
\ref{main} a lower bound $Q^{k,tt}(G_{M,N})\geq \Omega(p(N,M,k))$, 
where $p(N,M,k)= \mathrm{min}\left\{\frac{N}{M^2\cdot 
2^{(k+2)/M^{1/3}}},\frac{M-M^{1/3}}{(2M^{1/3}\log{M} 
+k+2)^2}\right\}$.  These two lower bounds of the nonadaptive 
query complexity will be proven in Sections \ref{sec:multiple-block} 
and \ref{sec:single-block}. Moreover, a large gap between 
$Q(G_{M,N})$ and $Q^{k,tt}(G_{M,N})$ will be used in Section 
\ref{sec:other-application} to separate adaptive and nonadaptive 
complexity classes. 

The aforementioned upper and lower bounds of the quantum 
nonadaptive query complexity of the multiple-block and 
single-block ordered search problems are summarized in Table 1.

\section{Query Complexity for Multiple Block Ordered Search}
\label{sec:multiple-block}

We demonstrate how to use an algorithmic argument to obtain a 
new query complexity lower bound for the multiple-block ordered 
search problem $G_{M,N}$. As a special case, the query complexity 
for the single-block ordered search problem will be discussed 
later in Section \ref{sec:single-block}. Now, we prove the 
following lower bound of $Q^{k,tt}(G_{M,N})$ as a main theorem. 

\begin{theorem}\label{main}
$Q^{k,tt}(G_{M,N})\geq\Omega\left(\mathrm{min}
\left\{ \frac{N}{ M^2\cdot 2^{(k+2)/M^{1/3}}}, 
\frac{M-M^{1/3}}{(2M^{1/3}\log M+k+2)^2}\right\}\right)$.
\end{theorem}

Theorem \ref{main} intuitively states that multiple-block ordered 
search requires a large number of nonadaptive queries even with 
the help of a relatively large amount of advice. To show the 
desired lower bound, we employ an algorithmic argument that revolves 
around the incompressibility of instances. 

We first prove a key proposition from which our main theorem 
follows immediately. For convenience, for any constant 
$\epsilon\in[0,1/2)$, we define $d(\epsilon)= 1/(2\epsilon)-1$ if 
$\epsilon>0$ and $d(\epsilon)=1$ otherwise. Letting $c$ be any 
constant satisfying $0<c<d(\epsilon)$, we further define 
$\epsilon'=(1+c)\epsilon$ and $C_\epsilon=(1-2\epsilon')^2/16$. 
For any string $y$ and any number $a\in[0,|y|]_{\integer}$, 
$\mathrm{First}_{a}(y)$ ($\mathrm{Last}_{a}(y)$, resp.) denotes 
the first (last, resp.) $a$-bit segment of $y$. Clearly, 
$y=\mathrm{First}_{|y|-a}(y)\mathrm{Last}_{a}(y)$.

To describe the key proposition, we need the notion of the 
{\em weight} of a query word and the function $C_{U,V}$. Fix 
$n,M\in\nat$ and $p\in[1,n]_{\integer}$ and set $N=2^n$. Consider 
the multiple-block problem $G_{M,N,p}$. Assume that a nonadaptive 
black-box quantum computer $(U,V)$ solves $G_{M,N,p}$ with error 
probability $\leq\epsilon$ with advice of length $k$ using $T$ 
nonadaptive queries. Let $s$ be any string of length $Mn$ and 
assume that $s=s_1s_2\cdots s_M$, where each $s_i$ is the $i$th 
block segment of $s$ with $|s_i|=n$. Let $f$ be its corresponding 
$k$-bit advice string. For any $i,j\in [1,M]_{\integer}$ and any 
$z\in\Sigma^n$, the {\em weight} of the query word $(j,z)$, 
denoted  $wt(i:j,z)$, is the sum of all the squared magnitudes of 
amplitudes of $\qubit{\vec{y}}\qubit{0^T}\qubit{\phi_{i,f,\vec{y} } }$ 
such that the list $\vec{y}=(y_1,\ldots,y_T)$ of query words 
contains $(j,z)$ in the prequery quantum state 
$U(\qubit{0}\qubit{0}\qubit{i,f}) 
=\sum_{\vec{y}}\qubit{\vec{y}}\qubit{0^T}\qubit{\phi_{i,f,\vec{y}}}$. 
Moreover, for each $i,j\in[1,M]_{\integer}$ and $z'\in\Sigma^{n-p}$, 
let $wt_p(i:j,z')$ be the sum of the values $wt(i:j,z)$ over 
all $z\in\Sigma^n$ satisfying $z'=\mathrm{First}_{n-p}(z)$. An 
index $i$ is called {\em good} if $wt_p(i:i,\mathrm{First}_{n-p}(s_i)) 
>C_\epsilon$. Any index that is not good is called {\em bad}. 
Let $l'_{s}$ denote the total number of good indices; \ie 
$l'_{s} = |\{i\in[1,M]_{\integer} \mid wt_p(i:i,\mathrm{First}_{n-p}(s_i))
>C_\epsilon \}|$. Note that $0\leq l'_{s}\leq M$. At length, 
the function $C_{U,V}$ is introduced in the following way:
\[
C_{U,V}(M,N,k,p,s,l)=
\left\{
\begin{array}{ll}
\frac{C_\epsilon N}{M^2 2^{p+1+(k+2)/l} }      
    & \mbox{if}\ \ l\leq l'_{s},\\
\frac{C_\epsilon (M-l)p^2}{( 2l\log{M} +k+2)^2} 
    & \mbox{if}\ \ l> l'_{s},
\end{array}
\right.
\]
where $l\geq 1$. The key proposition below relates to a 
relationship between the query complexity and the function 
$C_{U,V}$.  

\begin{proposition}\label{submain}
Let $\epsilon\in[0,1/2)$ and $c\in(0,d(\epsilon))$ and set 
$\epsilon'=(1+c)\epsilon$ and $C_\epsilon=(1-2\epsilon')^2/16$. 
Let $n,M\in\nat$ and $p\in[1,n]_{\integer}$ and set $N=2^n$. 
If a nonadaptive black-box quantum computer $(U,V)$ solves $G_{M,N,p}$ with error 
probability $\leq\epsilon$ by $T$ queries with
advice of length $k$, then  $T\geq \mathrm{max}_{1\leq l\leq M}
\min_{s\in\Sigma^{Mn}}\{ C_{U,V}(M,N,k,p,s,l)\}$.
\end{proposition}

Clearly, Theorem \ref{main} follows from Proposition \ref{submain} 
by setting $p=1$ and $l=M^{1/3}$ since 
$Q^{k,tt}(G_{M,N,1})\leq Q^{k,tt}(G_{M,N})$ for any constant 
$k$ in $\nat$. 

Now, we detail the proof of Proposition \ref{submain} by 
employing our algorithmic argument. 
Assume to the contrary that 
Proposition \ref{submain} fails. Let $(U,V)$ be a nonadaptive 
black-box quantum computer that solves $G_{M,N,p}$ with error 
probability $\leq \epsilon$ with $T$ nonadaptive queries using 
advice of length $k$. By our assumption, there exists a number 
$l\in[1,M]_{\integer}$ such that $T < C_{U,V}(M,N,k,p,s,l)$ for 
all strings $s$ in $\Sigma^{Mn}$. It follows by a simple 
calculation that, for each $s\in\Sigma^{Mn}$, 
\begin{eqnarray}\label{eq111}
l(2\log M -n +\log(T/C_\epsilon)+p +1) +k+2 
&<& 0\ \ \mbox{if }l\leq l'_{s},\\ \label{eq112}
2l\log M +k+2 -p\sqrt{C_\epsilon(M-l)/T} 
&<& 0\ \ \mbox{if }l>l'_{s}.
\end{eqnarray}

Our goal is to define a compression scheme $E$ working on all 
strings in $\Sigma^{Mn}$ such that (i) $E$ is one-to-one and 
(ii) $E$ is length-decreasing\footnote{A function $f$ from 
$\Sigma^n$ to $\Sigma^*$ is called {\em length-decreasing} if 
$|f(x)|<|x|$ for all $x\in\Sigma^n$.}. These two conditions 
clearly lead to a contradiction since any length-decreasing 
function from $\Sigma^{Mn}$ to $\Sigma^*$ cannot be one-to-one 
(by the pigeonhole principle). More precisely, we wish to define an 
``encoding'' of $s$, denoted $E(s)$. We first show that 
$|E(s)|<|s|$ using the definition of $E$. To show the one-to-oneness of $E$, 
we want to construct a 
``generic'' deterministic decoding algorithm that takes $E(s)$ 
and outputs $s$ for any string $s\in\Sigma^{Mn}$ because this decoding algorithm 
guarantees the uniqueness of the encoding $E(s)$ of $s$. Therefore, we obtain a 
contradiction, as requested, and complete the proof.

Let $s$ be any string in $\Sigma^{Mn}$ and let $f$ be its 
corresponding advice string of length $k$. We split our proof 
into the following two cases: (1) $l\leq l'_{s}$ and (2) 
$l> l'_{s}$. 

\ms
{\sf (Case 1: $l\leq l'_{s}$)}\hs{2} 
The desired encoding $E(s)$ contains the following four items: 
(i) the advice string $f$,  (ii) the $2l'_{s}\log{M}$-bit string 
encoding in double binary all the $l'_{s}$ good indices, (iii) a 
separator $01$, and (iv) all the strings $e(i)$ for each 
$i\in[1,M]_{\integer}$, where $e(i)$ is defined as follows. In 
case where $i$ is good, $e(i)$ is of the form 
$(k_i,\mathrm{Last}_{p}(s_i))$ with $k_i = 
|\{a\in\Sigma^{n-p}\mid wt_p(i:i,a)> 
C_\epsilon\ \mathrm{and}\ a < \mathrm{First}_{n-p}(s_i)  
\mbox{ (lexicographically)} \}|$. If $i$ is bad, then 
$e(i)=s_i$. These four items are placed in $E(s)$ in order 
from (i) to (iv). 

The following lemma shows that $\ceilings{\log (T/C_\epsilon)}$ 
bits are sufficient to encode $k_i$ in binary. 

\begin{lemma}\label{claim3}
For each good $i$, $k_i< T/C_\epsilon$. 
\end{lemma}

\begin{proof}
Let $i$ be any good index and define the set $A_i=\{a\in\Sigma^{n-p} 
\mid wt_p(i:i,a) > C_\epsilon \}$. Obviously, $|A_i|\geq k_i$. It suffices to 
show that $|A_i|<T/C_{\epsilon}$. Recall first that the weight 
$wt_p(i:i,a)$ represents the value $\sum_{z\in\Sigma^p}
\sum_{\vec{y}:(i,u)\in \vec{y}}\|\qubit{\phi_{i,f,\vec{y}}}\|^2$, 
where ``$(i,u)\in\vec{y}$'' means that the list $\vec{y}$ contains query 
word $(i,u)$ in $\Gamma_{M,N}$. Since each query list $\vec{y}$ 
contains at most $T$ query words,  
\[
\sum_{a\in\Sigma^{n-p}} wt_p(i:i,a) = \sum_{u\in \Sigma^{n}}
\sum_{\vec{y}:(i,u)\in \vec{y}}\|\qubit{\phi_{i,f,\vec{y}}}\|^2
 = \sum_{\vec{y}}\sum_{u:(i,u)\in \vec{y}} 
 \|\qubit{\phi_{i,f,\vec{y}}}\|^2 
\leq T \cdot \sum_{\vec{y}}\|\qubit{\phi_{i,f,\vec{y}}}\|^2 \leq  T.
\]
The last inequality comes from the fact that 
$\sum_{\vec{y}}\|\qubit{\phi_{i,f,\vec{y}}}\|^2 =1$.  
It thus follows that $T \geq \sum_{a\in\Sigma^{n-p}} 
wt_p(i:i,a)\geq \sum_{a\in A_i} wt_p(i:i,a) > C_\epsilon|A_i|$, 
which implies that $|A_i| < T/C_{\epsilon}$, as requested.  
\end{proof}

By Lemma \ref{claim3}, the representation of any pair 
$(k_i,\mathrm{Last}_{p}(s_i))$ 
requires at most $\ceilings{\log(T/C_{\epsilon})}+p$ bits. The 
total length of the encoding $E(s)$ is thus bounded above by: 
\begin{eqnarray*}
|E(s)| &\leq& k+ 2l'_{s}\log{M} + 2 
+l'_{s}(\log{(T/C_\epsilon)} +p +1 )+(M-l'_{s})n \\
&\leq& M n +l(2\log{M}-n +\log{(T/C_\epsilon)} +p +1 )+ k+2 \ <\ M n,
\end{eqnarray*}
where the second inequality is obtained from Eq.(\ref{eq111}) 
and our assumption $l\leq l'_{s}$, and the last inequality 
comes from Eq.(\ref{eq111}). Since $|s|=Mn$, it follows that 
$|E(s)|<|s|$.

We next show that the encoding $E(s)$ is uniquely determined 
from $s$. To show this, we give a deterministic decoding algorithm 
that extracts $s$ from $E(s)$ for all $s\in\Sigma^{Mn}$ with 
$l'_{s}\geq l$. The desired decoding algorithm $\AAA$ is 
described as follows.

\begin{quote}
{\sf Decoding Algorithm $\AAA$: For each $i\in[1,M]_{\integer}$, 
we compute $s_i$ in the following manner. First, check whether $i$ 
is good by examining item (ii) of $E(s)$.  If $i$ is bad, then 
find $e(i)=s_i$ directly from item (iv). The remaining case is 
that $i$ is good. Note that $wt_p(i:i,\mathrm{First}_{n-p}(s_i))
>C_\epsilon$ and  $e(i)=(k_i,\mathrm{Last}_{p}(s_i))$.  Define 
$A_i$ to be the set of all $(n-p)$-bit strings $a$ with 
$wt_p(i:i,a)>C_\epsilon$. Find the lexicographically $k_i+1$st 
string in $A_i$ by preparing the prequery state classically. 
Obviously, this string equals 
$\mathrm{First}_{n-p}(s_i)$ by the definition of $k_i$. Use 
$\mathrm{Last}_{p}(s_i)$ to obtain the desired string 
$s_{i}=\mathrm{First}_{n-p}(s_i)\mathrm{Last}_{p}(s_i)$. Finally, 
output the decoded string $s=s_1s_2\cdots s_M$.
}
\end{quote}

Since $\AAA$ does not involve the computation of $V$, it 
is easy to show that $\AAA$ correctly outputs $s$ from $E(s)$. 

\ms
{\sf (Case 2: $l>l'_{s}$)}\hs{2} 
Different from Case (1), the encoding $E(s)$ includes the 
following six items: (i) the advice string $f$, (ii) the 
$2l'_{s}\log{M}$-bit string that encodes in double binary all 
$l'_{s}$ good indices, (iii) a separator $01$, (iv) all the 
strings $s_{i}$ for each good index $i$,  (v) all the strings 
$\mathrm{First}_{n-p}(s_i)$ for each bad index $i$, and (vi) 
an additional string $r$ of length 
$\leq p((M-l'_{s})-\sqrt{C_{\epsilon}(M-l'_{s})/T})$, which 
will be defined later. These items are placed in $E(s)$ orderly 
from (i) to (vi).

We begin with the estimation of the length of $E(s)$. By summing 
up all the items of $E(s)$, we can upper-bound its length 
$|E(s)|$ by:
\begin{eqnarray*}
|E(s)| 
&\leq&  k + 2l'_{s}\log{M} + 2 +l'_{s} n +(M-l'_{s})(n-p) + p((M-l'_{s}) -
\sqrt{C_{\epsilon}(M-l'_{s})/T} ) \\
&\leq & Mn + 2l \log{M} +k+2 - p(\sqrt{C_{\epsilon}(M-l)/T}) 
\ <\ Mn,
\end{eqnarray*}
where the second inequality comes from our assumption 
$l>l'_{s}$ and the fact that the derivative of the function 
${\cal F}(z)=-p(\sqrt{C_\epsilon(M-z)/T})+ 2z \log{M}$ satisfies 
${\cal F}'(z)\geq 0$ for any $z\in(0,M)$ and the last inequality 
follows from Eq.(\ref{eq112}).  Therefore, we obtain the desired 
inequality $|E(s)|<|s|$.

We still remain to define the string $r$. To describe it,
we need to search for indices of light query weight. The 
following procedure, called the {\em lightly weighted step 
search} (abbreviated LWSS), selects a series of steps of light 
query weight. As we will show later, this series of steps are 
redundant and therefore, we can eliminate them, causing the 
compression of $s$. Let $m$ be the positive solution of the 
equation $(T/C_\epsilon) m^2 -\left(T/C_\epsilon -1\right)m 
-(M-l'_{s}) =0$. (In the case where $m$ is a non-integer, we need to
round it down.)

\begin{quote}
{\sf Procedure {\rm LWSS}: Let $R_1=\setempty$ and 
$L_1=\{ i\in[1,M]_{\integer} \mid \mbox{$i$ is bad} \}$. Repeat 
the following procedure by incrementing $i$ by one until $i=m$. 
At round $i$, choose the lexicographically smallest index $w_i$ 
in the difference $L_i-R_i$. Simulate $U$ deterministically on 
input $(w_i,f)$ to generate the prequery quantum state 
$\qubit{\gamma_f(w_i)}$. 
For each bad index $j\in[1,M]_{\integer}$, compute the weight 
$wt_p(w_i: j,\mathrm{First}_{n-p}(s_{j}))$ in $\qubit{\gamma_f(w_i)}$. 
Define $R_{i+1}=R_{i}\cup \{w_i\}$ and $L_{i+1}= L_{i}\cap 
\{ j\in[1,M]_{\integer} \mid wt_p(w_i: j,\mathrm{First}_{n-p}(s_j)) 
< C_{\epsilon}/m \}$. Finally, set $W=R_{m+1}$. Output all the 
elements in $W$.
}
\end{quote}

We can prove the following lemma regarding {\rm LWSS}. 

\begin{lemma}\label{claim1}
{\rm LWSS} produces a unique series of $m$ distinct indices $w_1,w_2,\ldots,w_{m}$ 
such that, for any pair $i,j\in[1,m]_{\integer}$, if $i<j$ then 
$wt_p(w_i:w_j,\mathrm{First}_{n-p}(s_{w_j}) )<C_\epsilon/m$.
\end{lemma}

\begin{proof}
First, we show that $|L_i|\geq M-l'_s -(T/C_\epsilon)m(i-1)$ by induction on 
$i$ with $1\leq i\leq m$. 
In the basis case where $i=1$, this is true because the number of bad indices is 
$M-l'_s$. For the induction step, we assume by our induction hypothesis that 
$|L_i|\geq M-l'_s -(T/C_\epsilon) m(i-1)$. 
For convenience, set $L'=\{\mu\in[1,M]_\integer \mid 
wt_p(w_i:\mu,\mathrm{First}_{n-p}(s_\mu))< C_\epsilon/m\}$. 
Now, we claim that $|L_1-L'|\leq (T/C_\epsilon) m$. 
{}From the definition of $L'$, it holds that, 
for any $\mu\not\in L'$,  $wt_p(w_i:\mu,\mathrm{First}_{n-p}(s_\mu))\geq 
C_\epsilon/m$. It thus follows that $|L_1 -L'|\cdot (C_\epsilon/m)
\leq \sum_{\mu\in L_1-L'}wt_p(w_i:\mu,\mathrm{First}_{n-p}(s_\mu)) \leq T$, 
where the last inequality comes from the fact that the total weight of query words 
must be at most $T$. Therefore, we obtain the claim $|L_1 -L'|\leq (T/C_\epsilon) m$. 
Using the fact that $L_{i+1}=L_{i}\cap L'$, $|L_{i+1}|=|L_i|-|L_i-L_{i+1}|\geq 
|L_{i}|-|L_1-L'|$, which is further bounded by: 
\[
 |L_{i}|-|L_1-L'| \geq M-l'_s-(T/C_\epsilon) m(i-1)-(T/C_\epsilon) m =M-l'_s 
-(T/C_\epsilon) m i.
\]
Therefore, $|L_{i+1}|\geq M-l'_s -(T/C_\epsilon)m i$. This completes the 
induction step.

To guarantee the existence of the $m$ indices $w_1,\ldots,w_m$, we want to 
show that $L_{m}\neq\setempty$. Since $m(m-1)\leq C_\epsilon(M-l'_s-m)/T$ 
(even though $m$ is rounded down), it follows that
$|L_{m}|\geq M-l'_s-(T/C_\epsilon) m(m-1) \geq m$. Because of $|R_{m}|=m-1$, 
$L_{m}-R_{m}$ cannot be empty. 
This implies that $w_{m}$ truly exists. Note that, by the definition of $L_j$, 
$L_{j}=\{\mu\in[1,M]_\integer \mid \forall i<j\: 
[wt_p(w_i:\mu,\mathrm{First}_{n-p}(s_{\mu}) ) < C_\epsilon/m]\}$. 
Procedure {\rm LWSS} clearly ensures that $w_j$ belongs to $L_j$. 
Hence, for any $i<j$, $wt_p(w_i:w_j,\mathrm{First}_{n-p}(s_{w_j}) ) < 
C_\epsilon/{m}$.
\end{proof}

The key element $r$ in item (vi) is defined as follows. Let 
$v_i$ be the lexicographically $i$th element in the set 
$\{j\in [1,M]_{\integer}\mid \mbox{$j$ is bad and }j\not\in W\}$ 
and let $r_i$ be the last $p$ bits of $s_{v_i}$. The string 
$r=r_1r_2\cdots r_{M-l'_{s}-m}$ constitutes item (vi). The 
length $|r|$ is clearly at most 
$p\left( M-l'_{s}-\sqrt{C_\epsilon(M-l'_{s})/T} \right)$ since 
\[
m=\frac{(T/C_\epsilon-1)+\sqrt{4(T/C_\epsilon)(M-l'_{s})
+(T/C_\epsilon-1)^2}}{2(T/C_\epsilon)} \geq  
\sqrt{C_\epsilon(M-l'_{s})/T}.
\] 

Next, we want to show the uniqueness of our encoding $E(s)$ by 
constructing its decoding algorithm. First, we check which 
index $i$ is good by simply examining item (ii). For any good 
index $i$, we immediately obtain $s_i$ from item (iv). When $i$ 
is bad, however, we obtain only $\mathrm{First}_{n-p}(s_i)$ from 
item (v).  To obtain the last $p$ bits of $s_i$, we need to exploit 
item (vi) of $E(s)$ and simulate $(U,V)$ in a deterministic fashion. 
Since we cannot use the oracle $G_{M,N,p}$, we need to substitute 
its true oracle answers with their approximated values. The desired 
decoding algorithm $\BB$ is given as follows.

\begin{quote}
{\sf Decoding Algorithm $\BB$:
1) For any good index $i\in[1,M]_{\integer}$, obtain $s_i$ 
directly from item (iv) of $E(s)$. For the other indices, run 
{\rm LWSS} to compute $W=\{w_1,\ldots,w_{m}\}$. For any bad 
index $e$ outside of $W$, item (vi) provides $\mathrm{Last}_{p}(s_e)$. 
Combining it with $\mathrm{First}_{n-p}(s_e)$ from item (v), 
we obtain $s_e$.  

2) Let $e$ be any bad index in $W$. First, obtain 
$\mathrm{First}_{n-p}(s_e)$ from item (v). The remaining part, 
$\mathrm{Last}_{p}(s_{e})$, is obtained as follows. Repeat the 
following procedure starting at round 1 up to $m$. At round $i$ 
($1\leq i\leq m$), assume that the last $p$ bits of 
$s_{w_1},\ldots,s_{w_{i-1}}$ have been already obtained. Simulate 
$U$ deterministically on input $(w_i,f)$ to generate the prequery 
quantum state $\qubit{\gamma_f(w_i)}= 
\sum_{\vec{y}}\qubit{\vec{y}}\qubit{0}\qubit{\phi_{w_i,f,\vec{y}}}$. 
Transform $\qubit{\gamma_f(w_i)}$ into the state 
$\qubit{\gamma_f(w_i)^r}=\sum_{\vec{y}}
\qubit{\vec{y}}\qubit{u_{1,\vec{y}}u_{2,\vec{y}}\cdots 
u_{T,\vec{y}}}\qubit{\phi_{w_i,f,\vec{y}}}$ using the string $r$, 
where each bit $u_{j,\vec{y}}$ is defined below. Choose any list 
$\vec{y}$ of query words. Note that $\vec{y}$ is of the form 
$(y_1,y_2,\ldots,y_{T})$. Let $j\in[1,T]_{\integer}$ and assume 
that $y_j$ is of the form $(a,vz)$, where $a\in [1,M]_{\integer}$, 
$v\in\Sigma^{n-p}$, and $z\in\Sigma^p$. For simplicity, write $u_j$ 
for $u_{j,\vec{y}}$.
 
a) If $v$ is lexicographically smaller (larger, rep.) than 
$\mathrm{First}_{n-p}(s_{a})$, then let $u_j=0$ ($u_j=1$, resp.). 
Next, assume $v =\mathrm{First}_{n-p}(s_{a})$. In the case where 
either $a$ is good or $a\not\in W$, obtain $\mathrm{Last}_p(s_{a})$ 
from items (iv) and (vi) and let $u_j=1$ if $z \geq 
\mathrm{Last}_p(s_{a})$ and let $u_j=0$ otherwise. 

b) Consider the case where $a$ is in $W$. Let $b$ be the index 
satisfying $a=w_{b}$. There are two cases to consider.

b-i) Assume that $b<i$. Note that $\mathrm{Last}_p(s_{a})$ has 
been obtained at an earlier round. Define $u_j=1$ if 
$z\geq \mathrm{Last}_p(s_{a})$ and $u_j=0$ otherwise. 

b-ii) If $b\geq i$, then set $u_j=0$. In particular, if $b>i$, 
then Lemma \ref{claim1} implies that  
$wt_p(w_i:w_b,\mathrm{First}_{n-p}(s_{w_b}) )$ $< C_{\epsilon}/m$ 
and $wt_p(w_i:w_i,\mathrm{First}_{n-p}(s_{w_i}) )\leq C_{\epsilon}$. 

3) Simulate $V$ deterministically on input $\qubit{\gamma_f(w_i)^{r}}$.  
Find its output that is obtained with probability $\geq 1/2$. 
Such a string must be $\mathrm{Last}_p(s_{w_i})$. With the known 
string $\mathrm{First}_{n-p}(s_i)$, this gives the entire string 
$s_i$, as required.

4) Output the decoded string $s=s_1s_2\cdots s_{M}$.
}
\end{quote}

We need to verify that the decoding algorithm $\BB$ correctly 
extracts $s$ from $E(s)$. If $i$ is good, then $s_i$ can be 
correctly obtained from item (iv). If $i$ is bad and not in $W$, 
then $\BB$ computes $\mathrm{Last}_{p}(s_i)$.  Henceforth, we 
assume that $i$ is bad and in $W$. Let $j$ be the index satisfying 
$i=w_j$. The operator $U$ on input $(w_{j},f)$ generates the 
prequery quantum state $\qubit{\gamma_f(w_{j})^{r}}$. We need 
to prove that our approximation of the true oracle answers 
from $G_{M,N,p}$ suffices for the correct simulation of $(U,V)$. 
Let $\qubit{\gamma_f(w_{j})^{G}}$ be the true postquery 
quantum state; namely, $\sum_{\vec{y}}
\qubit{\vec{y}}\qubit{G_{M,N,p}(y_1)\cdots G_{M,N,p}(y_{T})}
 \qubit{\phi_{w_{j},f,\vec{y}}}$, 
where $\vec{y}=(y_1,\ldots,y_T)$. 
Now, we claim that $\qubit{\gamma_f(w_{j})^{r}}$ is close to 
$\qubit{\gamma_f(w_{j})^{G}}$.

\begin{lemma}\label{claim4}
$\|\qubit{\gamma_f(w_{j})^{r}}-\qubit{\gamma_f(w_{j})^G}\| 
\leq 2\sqrt{C_{\epsilon}}$. 
\end{lemma}

\begin{proof}
Consider the set $A$ of all query lists $\vec{y}$ that, for a certain 
number $b\geq j$ and a certain string $z\in\Sigma^{p}$,
include a query word $(w_b,\mbox{First}_{n-p}(s_{w_b})z)$,  which was 
dealt with at Step (b-ii). The value $\|\gamma_f(w_j)^r-\gamma_f(w_j)^G\|^2$ 
is estimated as follows:
\begin{eqnarray*}
\lefteqn{\|\gamma_f(w_j)^r-\gamma_f(w_j)^G\|^2}\hs{5} \\
&=&
\left\|\sum_{\vec{y}\in A}\qubit{\vec{y}}
(\qubit{u_{1,\vec{y}},\ldots,u_{T,\vec{y}}}-
\qubit{G_{M,N,p}(y_1)\cdots G_{M,N,p}(y_{T})})
\qubit{\phi_{w_{j},f,\vec{y}}}\right\|^2\\
&\leq& \sum_{\vec{y}\in A}
\left(\|\qubit{\vec{y}}\qubit{u_{1,\vec{y}},\ldots,u_{T,\vec{y}}}
\qubit{\phi_{w_{j},f,\vec{y}}}\|^2
+\|\qubit{\vec{y}}\qubit{G_{M,N,p}(y_1) 
\cdots G_{M,N,p}(y_{T})}\qubit{\phi_{w_{j},f,\vec{y}}}\|^2\right),
\end{eqnarray*}
which equals $\sum_{\vec{y}\in A}
2\|\qubit{\phi_{w_{j},f,\vec{y}}}\|^2$. This term is further bounded by:
\begin{eqnarray*}
\sum_{\vec{y}\in A}
2\|\qubit{\phi_{w_{j},f,\vec{y}}}\|^2 
&\leq& 2 \sum_{b:b\geq j} 
wt_p(w_j:w_b,\mathrm{First}_{n-p}(s_{w_b}))\\
&\leq& 2(|W|C_\epsilon/m+C_\epsilon) \:\:=\:\: 4C_\epsilon.
\end{eqnarray*}
The second inequality follows from Lemma \ref{claim1} and the bound
$wt_p(w_j:w_j,\mathrm{First}_{n-p}(s_{w_j}))\leq C_\epsilon$. 
Therefore, we obtain $\|\qubit{\gamma_f(w_j)^r}-
\qubit{\gamma_f(w_j)^G}\|^2\leq 4C_{\epsilon}$, which yields the lemma.
\end{proof}

By Lemma \ref{claim4}, using the approximated oracle answer 
$\qubit{\gamma_f(w_j)^r}$ instead, 
the operator $V$ produces a wrong solution with probability 
at most $2\sqrt{C_{\epsilon}}+\epsilon\leq 1/2-\epsilon{c}$. 
In other words, $V$ outputs the correct string 
$\mathrm{Last}_{p}(s_{w_j})$ with probability at least 
$1/2+\epsilon{c}$. Since $\epsilon{c}>0$, the deterministic simulation of $V$ 
correctly provides us with the true outcome of $V$. This guarantees that 
$\BB$ correctly outputs $s$ from $E(s)$. This ends the discussion of Case (2).

\ms

Combining Cases (1) and (2), we conclude that $E$ is a 
length-decreasing one-to-one function from $\Sigma^{Mn}$ to 
$\Sigma^{*}$, contradicting the pigeonhole principle. This 
completes the proof of Proposition \ref{submain}. 

\section{Query Complexity for Single Block Ordered Search}
\label{sec:single-block}

The single-block ordered search problem $G_{1,N}$ has been 
extensively studied in the literature for the lower bound of 
its quantum adaptive query complexity. Upon quantum nonadaptive 
query computation, this section demonstrates a lower bound of the query 
complexity $Q^{k,tt}(G_{1,M})$ in the presence of advice. Our algorithmic 
argument again proves its usefulness. We 
follow the notations introduced in Section \ref{sec:multiple-block}. 

\begin{theorem}\label{ordered-nonadaptive}
Let $n\in\nat$ and set $N=2^n$. For any $\epsilon\in[0,1/2)$ 
and any $c\in(0,d(\epsilon))$, 
$Q_\epsilon^{k,tt}(G_{1,N})\geq C_{\epsilon}N/2^{2k+2}$, where 
$\epsilon'=(1+c)\epsilon$ and $C_{\epsilon}=(1-2\epsilon')^2/16$.  
\end{theorem}

\begin{proof} 
Letting $p=k+1$, it suffices to show that $Q_\epsilon^{k,tt}(G_{1,N,p})\geq
 C_\epsilon N/2^{2k+2}$. 
Henceforth, we consider the case where 
$k+1\leq n$ since, otherwise, the theorem trivially holds. 
Taking $T < C_\epsilon N/2^{2k+2}$, we assume that there is a 
nonadaptive black-box quantum computer $(U,V)$ that solves the problem
$G_{1,N,p}$ with error probability $\leq\epsilon$ with advice 
of length $k$ by $T$ nonadaptive queries. Fix any string 
$s$ of length $n$ and let $f$ be its corresponding advice string 
in $\Sigma^k$. For simplicity, write $w_p$ for 
$wt_p(1:1,\mathrm{First}_{n-p}(s))$. We aim at defining an encoding 
scheme $E$ that can be proven to be length-decreasing and one-to-one. 
We need to consider the following two cases separately: (1) $w_p>C_\epsilon$ 
and (2) $w_p\leq C_\epsilon$. 

\ms
{\sf (Case 1: $w_p>C_{\epsilon}$)} 
The encoding $E(s)$ consists of (i) the advice string
$f$, (ii) the $p$-bit string
$\mathrm{Last}_p(s)$, and (iii) the number $e=|\{a\in\Sigma^{n-p}\mid 
wt_p(1:1,a)>C_\epsilon \text{ and }
a <  \mathrm{First}_{n-p}(s)\}|$ in binary. 
Similar to Lemma \ref{claim3}, $e$ can be expressed with at most  
$\ceilings{\log(T/C_\epsilon)}$ bits. Thus, the coding length
$|E(s)|$ is bounded above by:
\[
|E(s)| \leq k+p+\log(T/C_\epsilon)+1 =2k+2+\log(T/C_\epsilon) 
< n=|s|.
\]
The deterministic decoding algorithm for $E(s)$ is given in a fashion 
similar to Case (1) in the proof of Proposition \ref{submain}.  

\ms
{\sf (Case 2: $w_p\leq C_{\epsilon}$)} 
In this case, the desired encoding $E(s)$ includes two items: (i) $f$ and 
(ii) $\mathrm{First}_{n-p}(s)$.  
The length of $E(s)$ equals $k+n-p$, which is obviously $n-1$ and is 
clearly less than $|s|$. In the following deterministic manner, we uniquely 
extract $s$ from $E(s)$. 
Simulate $U$ deterministically to generate query lists. Using the information 
$\mathrm{First}_{n-p}(s)$,  
we can determine the true oracle answer to any query word whose 
first $n-p$ bits are different from $\mathrm{First}_{n-p}(s)$. 
For any other query word, we can replace its true oracle answer by its estimation 
$0$. Such a replacement may increase the error probability of $V$. As in 
Case (2) of the proof of Proposition \ref{submain}, nonetheless, the probability 
that $V$ produces a wrong solution is bounded above by 
$2\sqrt{C_\epsilon}+\epsilon=1/2-\epsilon{c}$ since $w_p \leq C_\epsilon$. 
Hence, the simulation of $V$ in a deterministic manner helps find the right 
solution $s$. We therefore extract $s$ from $E(s)$ successfully. 

\ms

Cases (1) and (2) imply that $E$ is length-decreasing and also one-to-one. 
This obviously leads to a contradiction against the pigeonhole principle and 
we thus obtain the theorem. 
\end{proof}

For the special case where $k=0$ and $\epsilon=1/3$, Theorem 
\ref{ordered-nonadaptive} gives a lower bound 
$Q^{tt}(G_{1,N})\geq \Omega(N)$, which is {\em optimal} if we 
ignore its constant multiplicative factor since 
$Q^{tt}(G_{1,N})$ is at most $N-1$. 

As for the single-block ordered search problem, we can also 
employ a quantum adversary argument to prove its quantum 
nonadaptive query complexity. Particularly, using an inner product method 
of H{\o}yer et al. \cite{HNS02}, we again obtain a similar lower bound of 
$Q^{k,tt}(G_{1,N})$. 

\begin{proposition}\label{inner-product}
For any constant $\epsilon\in[0,1/2)$, $Q^{k,tt}_\epsilon(G_{1,N})
\geq (1-2\sqrt{\epsilon(1-\epsilon)})(N/2^k-1)$. 
\end{proposition}  

\begin{proof}
Assume that a nonadaptive black-box quantum computer $(U,V)$ needs 
$T$ queries 
to solve the problem $G_{1,N}$ with error probability $\leq\epsilon$ using 
advice of length $k$. For each step $s$, $\hat{s}$ denotes the input 
string $0^{s-1}1^{N-s+1}$ to the black-box quantum computer and let 
$f(s)$ denote the advice string that minimizes the number of queries 
used for solving $G_{1,M}$ on the input $\hat{s}$. 
For each $d\in\Sigma^k$, define the set $A_d=\{s\in[1,N]_\integer 
\mid f(s)=d \}$ of advice strings. 

Now, consider an advice string $a\in\Sigma^k$ whose cardinality is at 
least $N/2^k$. Write $b=|A_a|$ and assume that $A_a=\{s_1,s_2,\ldots,s_b\}$ 
with $s_1<s_2<\cdots<s_b$. Recall that $O_s$ denotes the unitary 
operator representing $s$. Given an input $\hat{s}$, the final quantum 
state of the black-box quantum computer $(U,V)$ is 
$\qubit{\psi_s}=VO_sU\qubit{a}$. 
For any indices $i,j\in[1,b]_{\integer}$, let $I(i,j)$ be the inner 
product between $\qubit{\psi_{s_i}}$ and $\qubit{\psi_{s_j}}$. 
We focus on the value $\zeta = \sum_{t=1}^{b-1}|I(t,t+1)|$. Note that 
our assumption yields an upper bound $|I(t,t+1)|\leq 2\sqrt{\epsilon(1-\epsilon)}$ 
due to \cite{Amb02,HNS02}. This gives an upper bound 
$\zeta \leq 2\sqrt{\epsilon(1-\epsilon)}(b-1)$. We next show a 
lower bound of $\zeta$.
The prequery state of the machine is 
$U\qubit{a}=\sum_{\vec{i},z}\alpha_{\vec{i}z}\qubit{\vec{i}}\qubit{0}\qubit{z}$, 
where $\vec{i}=(i_1,\ldots,i_T)$ corresponds to $T$ query words and $z$ 
represents work bits. Hence, $\zeta = \sum_{t=1}^{b-1} 
\sum_{\vec{i},z} |\alpha_{\vec{i},z}|^2
\prod_{j=1}^T \langle (\hat{s}_t)_{i_j}|(\hat{s}_{t+1})_{i_j}\rangle$. 
By the choice of $s_t$ and $s_{t+1}$, the inner product 
$\langle (\hat{s}_t)_{i_j}|(\hat{s}_{t+1})_{i_j}\rangle$ becomes $0$ 
if $i_j\in[s_t,s_{t+1}-1]_{\integer}$ and $1$ otherwise.  
The term $\zeta$ is then estimated as:
\begin{eqnarray*}
\zeta &=& 
\sum_{t=1}^{b-1}\sum_{i_1\not\in[s_t,s_{t+1}-1]_\integer}
                    \sum_{i_2\not\in[s_t,s_{t+1}-1]_\integer}\cdots
                    \sum_{i_T\not\in[s_t,s_{t+1}-1]_\integer}
|\alpha_{\vec{i},z}|^2\\
&\geq& \sum_{t=1}^{b-1}\sum_{\vec{i},z} |\alpha_{\vec{i},z}|^2
-\sum_{j=1}^T\left( \sum_{t=1}^{b-1}\sum_{u=0}^{s_{t+1}-s_t-1}
\sum_{\vec{i}[j,s_t+u],z}
|\alpha_{\vec{i}[j,s_t+u],z}|^2\right),
\end{eqnarray*} 
where $\vec{i}[j,w]$ denotes $(i_1,i_2,\ldots,i_{j-1},w,i_{j+1},\ldots,i_T)$. 
Since $\sum_{\vec{i},z}|\alpha_{\vec{i},z}|^2=1$, the above inequality 
implies a lower bound $\zeta\geq (b-1)-T$. 

The above two bounds of $\zeta$ derive the inequality 
$2\sqrt{\epsilon(1-\epsilon)}(b-1)\ge b-1-T$, which immediately implies 
$T\geq (1-2\sqrt{\epsilon(1-\epsilon)} )(b-1)\geq 
(1-2\sqrt{\epsilon(1-\epsilon)})(N/{2^k}-1)$.   
\end{proof}

We note that it is not clear whether the above inner product method can be 
extended to the multiple-block ordered search problem. 
 
\section{Other Applications of an Algorithmic Argument}
\label{sec:other-application}

We have shown in the previous sections how our algorithmic 
argument proves query complexity lower bounds. Hereafter, we 
apply our algorithmic argument to two notions of computational 
complexity theory: quantum truth-table reducibility and quantum 
truth-table autoreducibility. Particularly for quantum truth-table autoreductions, 
we describe our algorithmic argument using the notion of   space-bounded 
Kolmogorov complexity.

\subsection{Quantum Truth-Table Reducibility}

The first example is a nonadaptive oracle separation between 
$\p$ and $\bqp/\poly$. Earlier, Buhrman and van Dam \cite{BD99} 
and Yamakami \cite{Yam03} investigated quantum parallel query 
computations (\ie all query words are pre-determined before the 
first oracle query). It is shown in \cite{Yam03} that there 
exists an oracle relative to which polynomial-time classical 
adaptive query computation is more powerful than polynomial-time 
quantum parallel query computation. 

We have introduced a black-box quantum computer as a nonuniform model of 
computation. To describe nonadaptive $\bqp$-computations, we need a 
uniformity notion. For simplicity, we introduce a ``uniform'' model of 
quantum truth-table query computation by simply replacing a nonadaptive
black-box quantum computer $(U,V)$ with a pair $(M,N)$ of 
polynomial-time multi-tape well-formed quantum Turing 
machines\footnote{Alternatively, we can use a uniform family of 
polynomial-size quantum circuits. The equivalence of a multiple-tape QTM 
model and a quantum circuit model follows from \cite{NO02,Yam99,Yao93}.} 
(QTMs, in short). The notion of a QTM was introduced by Deutsch \cite{Deu85} 
and later reformulated by Bernstein and Vazirani \cite{BV97}. 
A $k$-tape QTM $M$ is a 6-tuple $(Q,\Sigma,\Gamma,q_0,q_f,\delta)$, where 
$Q$ is a finite set of inner states, $\Sigma$ is a finite alphabet, $\Gamma$ 
is a finite input tape alphabet, $q_0$ is the initial inner state in $Q$, $q_f$ is the 
final (or halting) inner state in $Q$, and $\delta$ is a quantum transition 
function dictating the behavior of the machine $M$. This  function $\delta$ 
induces the unitary operator acting on the Hilbert space spanned by the basis set 
consisting of all configurations of $M$. We assume the reader's 
familiarity with QTMs (e.g., see \cite{BV97,ON00,Yam99} for more details).

A pair $(M,N)$ of QTMs recognizes a language $L$ 
with oracle $A$ in the following fashion 
similar to a nonadaptive black-box quantum computer. The machine $M$ is 
equipped with at least two input tapes: one of which carries an original input 
and another does an advice string. On any input $x$, $M$ generates a prequery 
quantum state $\qubit{\phi} = 
\sum_{i_1,\ldots,i_T} \qubit{i_1,\ldots,i_T}\qubit{0^T}
\qubit{\psi_{i_1,\ldots,i_T}}$, which may depend on $x$. 
In a single step, the oracle $A$ answers all the queries by transforming 
$\qubit{\phi}$ to 
the postquery quantum state $\qubit{\phi'} = 
\sum_{i_1,\ldots,i_T}\qubit{i_1,\ldots,i_T} \qubit{A(i_1)\cdots A(i_T)}
\qubit{\psi_{i_1,\ldots,i_T}}$. 
Finally, $N$ begins with $\qubit{\phi'}$ as its initial superposition and 
eventually produces $L(x)$ in the output tape with probability $\geq 2/3$. 
For our convenience sake, we henceforth call such a pair $(M,N)$ a {\em 
truth-table query QTM}. The relativized complexity class $\bqp_{tt}^A$ relative 
to oracle $A$ is then defined as the 
collection of all sets recognized by polynomial-time truth-table query
QTMs in the aforementioned manner using $A$ as an oracle. The class 
$\bqp_{tt}^A/\mathrm{poly}$ uses, in addition, polynomial advice. 
Note that $\bqp_{tt}^{A}\subseteq \bqp^A$ for any oracle set $A$.

Applying the result of Section \ref{sec:multiple-block}, we 
can show the following theorem. For its proof, we fix an effective enumeration 
$\{p_i\}_{i\in\nat^{+}}$ of all polynomials and an effective enumeration 
$\{(M_i,N_i)\}_{i\in\nat^{+}}$ of such polynomial-time truth-table query 
QTMs, each of which $(M_i,N_i)$ runs in time at most $p_i(n)$ on any input 
of length $n$. Moreover, for any set $A$ and any number $n\in\nat$, 
the notation $A[n]$ denotes the $2^n$-bit string 
$A(\mathrm{bin}_n(1))A(\mathrm{bin}_n(2))\cdots A(\mathrm{bin}_n(2^n))$.

\begin{theorem}\label{qtt-vs-Turing}
There is a recursive oracle $A$ such that 
$\p^A\nsubseteq \bqp_{tt}^A/\mathrm{poly}$. 
\end{theorem}

\begin{proof}
We first define $\AAA$ as the collection of all oracles $A$ such that, for every 
$n\in\nat$, there exist $2^n$ steps 
$s_1,s_2,\ldots,s_{2^n}\in[1,2^{3n}]_{\integer}$ satisfying 
$A[4n]=\hat{s}_1\hat{s}_2\cdots \hat{s}_{2^n}$, where each $\hat{s}_i$ 
denotes $0^{s_i-1}1^{2^{3n}-s_i+1}$. Using any oracle $A$ drawn from $\AAA$, 
we define the oracle-dependent set $L^A$ to be the collection of all strings 
of the form $\mathrm{bin}_{n}(i)$, where $n\in\nat$ and $i\in[1,2^n]_\integer$, 
such that $A[4n]=\hat{s}_1\hat{s}_2\cdots \hat{s}_{2^n}$ 
for certain $2^n$ steps $s_1,s_2,\ldots,s_{2^n}\in[1,2^{3n}]_{\integer}$ and 
$s_i\equiv1\:\:(\mbox{mod }2)$. Given any oracle $A\in\AAA$ and any 
input $\mathrm{bin}_{n}(i)$, 
we can easily find the step $s_i$ deterministically by binary search over the set 
$A[4n]$ in polynomial time. Note that binary search requires adaptive queries to 
$A$. It thus follows immediately that $L^A$ belongs to $\p^A$ for an arbitrary 
oracle $A$ in $\AAA$. 

To prove the theorem, we want to construct a special oracle $A$ in $\AAA$ 
that places $L^A$ outside of $\bqp_{tt}/\mathrm{poly}$ by diagonalizing 
against all polynomial-time truth-table query QTMs. Such an oracle $A$ will 
contain strings of certain lengths in $\Lambda$, where 
$\Lambda=\{ n_j \}_{j\in\nat}$ is any fixed subset of $\nat$ satisfying 
that $n_{j+1}>p_{j}(n_{j})$ for any number $j$ in $\nat$. By stages, we 
build the desired set $A=\bigcup_{i\in\nat}A_i$. We set $A_0=\setempty$ at 
stage $0$. At stage $j\geq1$, we focus our attention on the $j$th machine 
$(M_j,N_j)$ and henceforth let $n=n_j$ for simplicity. Intuitively,  
Proposition \ref{submain} implies, by taking $p=1$, $l=2^{n/2}$ and $k=2^{n/2}$, 
that any ``truth-table query QTM'' solving the multiple-block ordered search problem 
$G_{2^n,2^{3n},1}$ requires at least $2^{n/3}$ nonadaptive queries even with the use 
of advice of length $2^{n/2}$. In other words, there exist a block number 
$i\in[1,2^n]_{\integer}$ and a series of $2^n$ steps $s_1,s_2,\ldots, s_{2^{n}}
\in[1,2^{3n}]_{\integer}$ such that, given instance 
$(i,\hat{s}_1\cdots \hat{s}_{2^n})\in [1,2^n]_{\integer}\times\Sigma^{2^{4n}}$, 
$(M_j,N_j)$ needs $2^{n/3}$ nonadaptive queries on input 
$\mathrm{bin}_{n}(i)$ to compute the value $s_i\:\text{mod}\:2$ with high 
success probability even with the help of any advice string of length $2^{n/2}$. 
Choose the minimal such series $(s_1,s_2,\ldots, s_{2^{n}})$ and define $A_j$ to 
satisfy $A_j[4n]=\hat{s}_1\cdots \hat{s}_{2^n}$. 
This intuitive argument, nevertheless, ignores the fact that 
$M_j$ may make queries of words of length less than or greater than $4n$. To 
deal with such queries, we need to modify the proof of Proposition \ref{submain} 
in the way described below. Since $n_{j+1}>p_j(n_j)$, we can answer $0$ to all 
the queries of length $>n$. Notice that any future stage will not affect the 
machine's behavior on input $\mathrm{bin}_n(i)$. When $M_j$ queries a 
word $y$ of length $<n$, we deterministically re-construct the oracle 
$A\cap\Sigma^{\leq n-1}$ and compute the oracle answer $A(y)$ using 
$A\cap\Sigma^{\leq n-1}$. This modification adds only an extra additive constant 
term to the encoding size given in the proof of Proposition \ref{submain}. Hence, the 
main assertion of Proposition \ref{submain} is still valid and 
$(M_j,N_j)$ cannot recognize $L^{A\cap\Sigma^{\leq n}}$. 

The desired set $A=\bigcup_{j\in\nat}A_j$ clearly belongs to $\AAA$. Our 
construction further guarantees that $L^A$ is not in $\bqp_{tt}/\mathrm{poly}$. 
Moreover, $A$ can be recursive since every $(M_i,N_i)$ is a uniform model 
and the proof of Proposition \ref{submain} is constructive.
\end{proof}

\subsection{Quantum Truth-Table Autoreducibility}
\label{sec:autoreduction}

As the second example, we focus our interest on the notion 
of autoreducible sets. After Trakhtenbrot \cite{Tra70} 
brought in the notion of autoreduction in recursion theory, 
the autoreducible sets have been studied in, \eg program 
verification theory. In connection to the program checking 
of Blum and Kannan \cite{BK89}, Yao \cite{Yao90} is the first 
to study $\bpp$-autoreducible sets under the name 
``coherent sets,'' where a set $A$ is {\em $\bpp$-autoreducible} 
if there is a polynomial-time oracle probabilistic Turing 
machine (PTM, in short) with oracle $A$ which determines 
whether any given input $x$ belongs to $A$ with probability $\geq 2/3$  
without querying the query word $x$ itself. Let $\bpp\mbox{-}\mathrm{AUTO}$ 
denote the class of all $\bpp$-autoreducible sets. Yao showed 
that $\mathrm{DSPACE}(2^{n^{\log\log n}})\nsubseteq
\bpp\mbox{-}\mathrm{AUTO}$. Later, Beigel and Feigenbaum 
\cite{BF92} presented a set in $\mathrm{ESPACE}$ which is 
not $\bpp$-autoreducible even with polynomial advice. If only 
nonadaptive queries are allowed in the definition of a $\bpp$-autoreducible 
set, it is specifically called 
{\em nonadaptively $\bpp$-autoreducible}. Feigenbaum, Fortnow, 
Laplante, and Naik \cite{FFLN98} gave an adaptively 
$\bpp$-autoreducible set which is not nonadaptively 
$\bpp$-autoreducible even with polynomial advice.  

We consider a quantum analogue of nonadaptively 
$\bpp$-autoreducible sets, called {\em $\bqp$-tt-autoreducible} 
sets, where ``tt'' is used to emphasize the nature of 
``truth-table'' queries rather than parallel queries. Formally, 
we obtain a $\bqp$-tt-autoreducible set by replacing a 
polynomial-time PTM in the above definition by a  
polynomial-time truth-table query QTM $(M,N)$, provided that any prequery quantum 
state produced by $M$ on each input $x$ does not include the 
query word $x$ with nonzero amplitude. Let 
$\bqp_{tt}/\mathrm{poly\mbox{-}AUTO}$ be the class of 
all $\bqp$-tt-autoreducible sets with polynomial advice. 

We prove the following separation, which extends the aforementioned result of 
Feigenbaum \etalc~\cite{FFLN98}. 

\begin{theorem}\label{pauto-vs-bqpttauto}
$\bpp\mbox{-}\mathrm{AUTO} \nsubseteq \bqp_{tt}/\mathrm{poly}
\mbox{-}\mathrm{AUTO}$. 
\end{theorem} 

\begin{proof}
{}From the proof of Theorem \ref{qtt-vs-Turing}, we recall the collection 
$\AAA$ of oracles. In addition, we introduce the oracle-dependent set $K_A$ for 
each oracle $A\in\AAA$ as follows. For any number $n\in\nat$ and any 
two indices $i\in[1,2^n]_{\integer}$ and $j\in[1,2^{3n}]_{\integer}$, write $(i,j)_n$ 
to denote $(i-1)\cdot2^{3n}+j$, which indicates the $j$th location in the $i$th block. 
For any string $x$ of the form $\mathrm{bin}_{4n}((i,j)_n)$, where $n\in\nat$, 
$i\in[1,2^n]_{\integer}$, and $j\in[1,2^{3n}]_{\integer}$, $x$ is in $K_A$ if 
either (i) $j=1$ and $s_i\equiv1\:(\mbox{mod }2)$ or (ii) $j\neq 1$ and $A(x)=1$. 

We first claim that $K_A$ belongs to $\bpp\mbox{-}\mathrm{AUTO}$ for any 
choice of $A$ from $\AAA$. For any nonempty string $x\in\Sigma^*$, let $x^{+}$ 
($x^{-}$, resp.) denote the lexicographic successor (predecessor, resp.) of $x$. Let $x$ be 
an arbitrary string of the form $\mathrm{bin}_{4n}((i,j)_n)$ for a certain 
choice of $n\in\nat$, $i\in[1,2^n]_{\integer}$, and $j\in[1,2^{3n}]_{\integer}$.
When $j=2^{3n}$, we immediately output $1$. If $j$ is in $[2,2^{3n}-1]_{\integer}$, then 
we first make two queries $x^{-}$ and $x^{+}$ to the set $K_A$ given as an oracle. If 
$K_A(x^{-})=K_A(x^{+})$, then we output $K_A(x^{+})$. On the contrary, if 
$K_A(x^{-})<K_A(x^{+})$, then we make an additional query $\mathrm{bin}_{4n}((i,1)_n)$ to 
$K_A$ and output its oracle answer. In the last case where $j=1$, we perform binary 
search over the set $K_A[4n]$ to determine the $i$th step $s_i$ and output the value 
$s_i\:\mbox{mod}\:2$, which equals $K_A(x)$.

Next, we show that, for a certain choice of $A$ from $\AAA$, $K_A$ does not 
belong to $\bqp_{tt}/\mathrm{poly\mbox{-}AUTO}$. 
We wish to construct such an $A$ by stages. At each stage, we choose a new polynomial-time 
truth-table query QTM $(M,N)$ and also take an input $(i,1)_n$ which is large enough for the 
diagonalization below. Assume that $(M,N)$ computes $K_A((i,1)_n)$ 
($=s_i\:\mbox{mod }2$) with high probability. First, recall the definition of the 
weight $wt_1(i:i',d)$, which denotes the sum of the squared magnitudes 
of the amplitudes of all vectors $\qubit{\vec{y}}\qubit{0}\qubit{\phi_{i,\vec{y}}}$ 
in the prequery state $M\qubit{x}$ such that $\vec{y}$ contains either 
query word $\mathrm{bin}_{4n}((i',2d-1)_n)$ or 
$\mathrm{bin}_{4n}((i',2d)_n)$ for a certain number 
$d\in[1,2^{3n-1}]_{\integer}$. Since our oracle is $K_A$ instead of $A$, 
we need to modify $wt_1(i:i',d)$ by adding the squared magnitudes of the 
amplitudes of all states $\qubit{\vec{y}}\qubit{0}\qubit{\phi_{i,\vec{y}}}$ in which $\vec{y}$ 
contains $(i',1)_n$. A proof similar to that of Proposition \ref{submain} together with a 
slight modification given in the proof of Theorem \ref{qtt-vs-Turing} works to prove a 
superpolynomial lower bound of the nonadaptive query complexity on the computation 
of $(M,N)$. Therefore, there exists an input $(i,1)_n$ on which $(M,N)$ fails to 
compute $K_A((i,1)_n)$ with high probability.   
\end{proof}

Beigel and Feigenbaum \cite{BF92} showed that 
$\mathrm{ESPACE}\nsubseteq \bpp/\mathrm{poly}\mbox{-}\mathrm{AUTO}$. 
It is also proven in  \cite{NY04} that 
$\mathrm{ESPACE}\nsubseteq \bqp/\mathrm{poly}$. In addition to 
these results, we show the existence of a set in $\mathrm{ESPACE}$ 
which is not $\bqp$-tt-autoreducible even with polynomial advice. 
To show this, we apply our algorithmic argument to a 
space-bounded compression algorithm. 

\begin{theorem}\label{maintheorem}
$\mathrm{ESPACE}\nsubseteq \bqp_{tt}/\mathrm{poly}\mbox{-}\mathrm{AUTO}$.
\end{theorem}

Note that Theorem \ref{maintheorem} is incomparable to the 
aforementioned results in \cite{BF92,NY04}. 
To prove Theorem \ref{maintheorem}, we first prove a key lemma on the 
space-bounded Kolmogorov complexity  of any set in 
$\bqp_{tt}/\mathrm{poly\mbox{-}AUTO}$. 
We need to fix a universal (deterministic) Turing machine $\MM_{U}$ in 
the rest of this section. Let $q$ be any function mapping from $\nat$ to $\nat$.
The {\em conditional $q$-space bounded Kolmogorov complexity 
of $x$ conditional to $s$}, denoted $\mathrm{C}^{q}(x|s)$, 
is the minimal length of any binary string $w$ such that, 
on input $(w,s)$, $\MM_{U}$ produces $x$ in its output tape using space 
at most $q(|x|+|s|)$ (see, \eg \cite{LV97} for more details).
We now present the following technical lemma. 

\begin{lemma}\label{mainlemma}
Let $A$ be any set in $\bqp_{tt}/\mathrm{poly}\mbox{-}\mathrm{AUTO}$ with a 
polynomial advice function $h$ 
such that $A\subseteq\bigcup_{n\in\mathrm{Tower2}}\Sigma^{n}$ and that 
the number of queries 
to $A$ is $t(n)$ on any input of length $n$ for any $n\in\nat$. There exist a 
polynomial $q$ and a constant $c\geq0$ such that, 
for any sufficiently large number $n\in\mathrm{Tower2}$, the space-bounded 
Kolmogorov complexity $\mathrm{C}^{q}(A[n]|h(n))$ is bounded above by
$2^n-m+2n+2\log{n}+c$, 
where $m$ is the positive solution of $288t(n)m^2-(288t(n)-1)m-2^n=0$.
\end{lemma}

\begin{proof}
Let $A$ be any set in $\bqp_{tt}/\mathrm{poly}\mbox{-}\mathrm{AUTO}$ with a 
polynomial advice function $h$. 
There exist a polynomial $t$ and a polynomial-time truth-table query QTM $(M,N)$ 
such that (i) on input $(x,h(|x|))$, $M$ outputs the prequery state   
$\qubit{\gamma}$ $=$ 
$\sum_{\vec{y}}\qubit{\vec{y}}\qubit{0^{t(|x|)}}\qubit{\phi_{\vec{y}}}$, (ii)
$wt(x:x)=0$, and (iii) $N(\qubit{\gamma^{A}})$  
outputs $A(x)$ with error probability at most $1/3$, where 
$\qubit{\gamma^{A}}=\sum_{\vec{y}}\qubit{\vec{y}}
\qubit{A(y_1)A(y_2)\cdots A(y_{t(|x|)})}\qubit{\phi_{\vec{y}}}$ for 
$\vec{y}=(y_1,y_2,\ldots,y_{t(|x|)})$. 
For any pair $x,z\in\Sigma^n$, we write $wt(x:z)$ for the sum 
of all squared magnitudes $\|\qubit{\vec{y}}\qubit{ 0^{t(|x|)} }
\qubit{\phi_{\vec{y}}}\|^2$ over all query lists $\vec{y}$ that contain $z$. 
Take any sufficiently large integer $n$ and fix it.

Similar to the proof of Theorem \ref{qtt-vs-Turing}, we need to deal with $M$'s 
query words by simulating $M$'s computation in a deterministic manner. 
Note that, since $M$'s running time is polynomially bounded, for any sufficiently 
large number $n$, $M$ cannot make any query of length $\geq 2^n$. Moreover, 
when $M$ queries words of length between $\log{n}+1$ and 
$n-1$, since $A\subseteq\bigcup_{n\in\mathrm{Tower2}}\Sigma^n$, we 
know that the oracle answers negatively.  Only query words of length 
$\leq \log{n}$ need our attention.

Consider the following deterministic procedure $\mathrm{LWSS2}$.  
For convenience, abbreviate $t(n)$ as $t$ in the rest of the proof. 
Let $m$ be the positive solution of $288tm^2-(288t-1)m-2^n=0$. Note that 
$\sqrt{2^{n}/576t}\leq m\leq 2^n$.
\begin{quote}
{\sf Procedure $\mathrm{LWSS2}$: Initially, set $R_1=\setempty$ and 
$L_1=\Sigma^{n}$. 
Repeat the following procedure by incrementing $i$ by one while $i\leq m$. 
At round $i\in[1,m]_{\integer}$, choose the lexicographically smallest string $w_i$ in 
$L_{i}-R_{i}$. 
Simulate $M$ deterministically on input $(w_i,h(n))$ to generate 
$\qubit{\gamma_i}= \sum_{\vec{y}}\qubit{\vec{y}}\qubit{0^{t(|x|)}}\qubit{\phi_{\vec{y}}}$. 
For each query word $y$, compute its weight $wt(w_i:y)$ in $\qubit{\gamma_i}$. 
Define $R_{i+1}=R_{i}\cup \{w_i\}$ and $L_{i+1}=L_{i}\cap \{y\in\Sigma^n 
\mid wt(w_i:y)< \frac{1}{288m}\}$. 
Finally, set $W=R_{m+1}$. Output all the elements in $W$.}   
\end{quote}
Note that procedure $\mathrm{LWSS2}$ uses space $2^{O(n)}$ 
since it deterministically simulates all computation paths of $M$ one by one 
and computes the weights of query words along these paths and stores the 
contents of $L_{i}$ and $R_{i}$. 
The following lemma can be proven similar to Lemma \ref{claim1}. 

\begin{lemma}\label{claim2}
$\mathrm{LWSS2}$ produces a unique series of $m$ distinct strings $w_1,w_2$, 
$\ldots$,$w_{m}$ 
such that, for any pair $i,j\in[1,m]_{\integer}$, 
$j>i$ implies $wt(w_{i}:w_{j})<\frac{1}{288m}$. 
\end{lemma}

For each $i\in[1,2^n-m]_{\integer}$, let $v_i$ be the 
lexicographically $i$th element 
in the difference set $\Sigma^n-W$ and set $r=A(v_1)A(v_2)\cdots A(v_{2^n-m})$. 
Recall the notation $A[n]=A(\mathrm{bin}_{n}(1))\cdots 
A(\mathrm{bin}_{n}(2^{n}))$ and define $z=A[0]A[1]\cdots A[\log{n}]$, which 
contains all the information on $A\cap\Sigma^{\leq \log{n}}$. Note that 
$|z|=2n-1$ if $n\geq1$.
 
Consider the following deterministic algorithm ${\cal C}$ that produces $A[n]$ 
on input $h(n)$ and extra information on $n$, $r$, and $z$.
\begin{quote}
{\sf Algorithm ${\cal C}$: 
1) On input $h(n)$, retrieve the hardwired number $n$ and the 
strings $r=r_1r_2\cdots r_{2^n-m}$ and $z=z_1z_2\cdots z_{\log{n}}$, where each 
$r_i$ is in $\{0,1\}$ and each $z_i$ is in $\Sigma^i$. First, run 
$\mathrm{LWSS2}$ to obtain $W=\{w_1,w_2,\ldots,w_m\}$. 

2) Choose every string $y$ in $\Sigma^n-W$ lexicographically one by one 
and find the number $k$ such that $y=v_k$. Clearly, $r_k$ matches $A(y)$. 

3) In this phase, we compute all the values $A(w_i)$ for $i\in[1,m]_{\integer}$. 
Repeat the following procedure. At round $i$ ($1\leq i\leq m$), 
assume that the $i-1$ values $A(w_1),A(w_2),\ldots,A(w_{i-1})$ have been 
already computed.
Simulate $M$ on input $(w_i,h(n))$ deterministically to generate 
$\qubit{\gamma_i}= \sum_{\vec{y}}\qubit{\vec{y}}\qubit{0^t}\qubit{\phi_{\vec{y}}}$. 
Using the extra information $r$, generate the 
vector $\qubit{\gamma^{r}}=
\sum_{\vec{y}}\qubit{\vec{y}}\qubit{u_1u_2\cdots u_t}\qubit{\phi_{\vec{y}}}$ 
as follows. 
Let $\vec{y}=(y_1,y_2,\ldots,y_t)$ be any query list in $\qubit{\gamma_i}$. 
For each $j\in[1,t]_{\integer}$, we determine the value $u_j$ as follows. 

i) In the case where $y_j\in W$, first find the index $k$ such that $y_j=w_k$. 
If $k<i$, let $u_j$ be the value $A(w_k)$ and otherwise, set $u_j=0$. 
Note that $wt(w_i:w_k)<\frac{1}{288m}$ by Lemma \ref{claim2}. 

ii) If $y_j\in\Sigma^n-W$, then we choose $k$ such that 
$y_j=v_k$ and define $u_j=r_k$. 

iii) Assume that $|y_j|\neq n$. As noted before, since $|y_j|<2^n$, if 
$|y_j|>n$ then let $u_j=0$. If $\log{n}<|y_j|<n$, then let $u_j=0$. Assume that 
$|y_j|\leq\log{n}$. Assume that $y_j$ is the lexicographically $k$th string in 
$\Sigma^{\leq \log{n}}$. In this case, let $u_j$ be the $k$th bit of $z$.

Finally, simulate $N$ on input $\qubit{\gamma^{r}}$ deterministically. 
There exists the unique output $z$ that is obtained by $N$ with probability $>1/2$. 
This $z$ must be $A(w_i)$. 

4) Finally, output the $2^n$-bit string 
$A(\mathrm{bin}_n(1))A(\mathrm{bin}_n(2))\cdots A(\mathrm{bin}_n(2^n))$ and halt. 
}
\end{quote}
Algorithm $\CC$ uses $2^{O(n)}$ space on input $(h(n),n,r,z)$ since {\rm LWSS2} 
requires only $2^{O(n)}$ space. Hence, we can choose an appropriate polynomial 
$q$ satisfying that $\CC$ runs using space at most $q(2^{n})$ for any $n\in\nat$. 

Now, we wish to prove that ${\cal C}$ correctly produces $A[n]$. Let 
$i\in[1,2^n]_{\integer}$. 
If $\mathrm{bin}_n(i)\not\in W$, then $A(\mathrm{bin}_n(i))$ is directly 
obtained from $r$. 
Assume that $\mathrm{bin}_n(i)\in W$ and let $j$ satisfy $w_{j}=\mathrm{bin}_n(i)$. 
On input $(w_{j},h(n))$, $\CC$  generates the quantum state $\qubit{\gamma^{r}}$. 
By a calculation similar to Lemma \ref{claim4}, it follows that 
$\|\qubit{\gamma^{r}}-\qubit{\gamma^{A}}\|^2\leq 2\cdot 
|W|\cdot \frac{1}{288m}= 1/144$. Thus, the error probability of $N$ 
is at most $\sqrt{1/144}+1/3=5/12$. 
This implies that the output bit obtained by $N$ with probability 
$>1/2$ matches the true value $A(\mathrm{bin}_n(i))$.                     
Therefore, $\CC$ correctly outputs $A(\mathrm{bin}_n(i))$ since ${\cal C}$ is 
deterministic. 

Recall that $\CC$ uses the hardwired information $(n,r,z)$, which is given as the 
concatenation of the following four items: (i) the string expression of $n$ in double 
binary, (ii) a separator $01$ , (iii) the string $r$, and (iv) the string $z$. 
Simulating $\CC$ on the universal machine $\MM_{U}$, we obtain  
$\mathrm{C}^{q}(A[n]|h(n))\leq 2\log{n}+ |r| +|z| +c \leq 2^n-m+2n+2\log{n}+c$, 
where $c$ is a certain nonnegative constant independent of the choice of $n$. 
This completes the proof.
\end{proof}

Using Lemma \ref{mainlemma} and a diagonalization method, we finally prove 
the desired theorem. In the following proof, we assume that a 
standard paring function $\pair{\hs{1},\hs{1}}$ from $\nat\times\nat$ to $\nat$ 
with $\pair{0,0}=0$. 

\begin{proofof}{Theorem \ref{maintheorem}}
We want to construct a set $A$ in $\mathrm{ESPACE}$ by stages. To 
simplify the proof, we fix the set $\{p_i\}_{i\in\nat}$ of polynomials 
that satisfy $p_i(n)=n^{i+1}+i$ for all $n\in\nat$. Note that, for any 
two indices $i,j\in\nat$ and any string $y\in\Sigma^*$ (i)
$\mathrm{C}^{p_i}(y|z)\geq \mathrm{C}^{p_{i+1}}(y|z)$ for any string 
$z$ and (ii) $\min_{z:|z|\leq p_{j}(n)}\mathrm{C}^{p_i}(y|z)\geq 
\min_{z':|z'|\leq p_{j+1}(n)}\mathrm{C}^{p_i}(y|z')$ for any 
number $n\in\nat$. Initially, set $n_0=1$ and $A_0=\setempty$ at stage $0$. 
At stage $k=\pair{i,j}\geq1$, take the minimal number $n_k$ in 
$\mathrm{Tower2}$ such that $n_k>n_{k-1}$, $2^{n_k/4}\geq p_j(n_k)+1$, and $2^{n_k/3} > 6n_k$.  
Take also the lexicographically minimal $2^{n_k}$-bit string $y$ such that 
$\mathrm{C}^{p_i}(y|z)\geq 2^{n_k}-2^{n_k/4}$ for all $z\in\Sigma^{\leq p_j(n_k)}$. 
We then define $A_k$ so that $A_k[n_k]=y$. Such a $y$ exists because of the 
following lemma.

\begin{lemma}\label{claim5}
There exists a string $y\in\Sigma^{2^{n_k}}$ such that 
$\mathrm{C}^{p_l}(y|z)\geq 2^{n_k}-2^{n_k/4}$ for every $l\leq i$, $m\leq j$, and 
$z\in\Sigma^{\leq p_m(n_k)}$.
\end{lemma}

\begin{proof}
By our choice of polynomials, it suffices to prove the lemma for $l=i$ and $m=j$. 
Let $g=2^{n_k}-p_j(n_k)-1$. The definition of $n_k$ implies that $g\geq 2^{n_k}-2^{n_k/4}$. Now, we assume otherwise that, for every 
$y\in\Sigma^{2^{n_k}}$, there exist a string $z\in\Sigma^{\leq p_j(n_k)}$ 
and a program $w\in\Sigma^{<g}$ such that $\MM_{U}(w,z)$ outputs $y$ 
using space at most $p_i(|y|+|z|)$. For each $y$, define $B_y$ as the 
collection of all pairs $(w,z)\in \Sigma^{<g}\times\Sigma^{\leq p_j(n_k)}$ such that 
$\MM_{U}(w,z)$ outputs $y$ using space $\leq p_i(|y|+|z|)$.  Since 
$|B_y|\geq1$ for every $y\in\Sigma^{2^{n_k}}$, we obtain the 
inequality $2^{2^{n_k}}\leq \sum_{y:|y|=2^{n_k}}|B_y|$. Note that, for any 
pair $y,y'\in\Sigma^{2^{n_k}}$, $B_y\cap B_{y'}=\setempty$ if $y\neq y'$. 
A simple estimation thus shows that:
\[
\sum_{y:|y|=2^{n_k}}|B_y| = 
\left|\bigcup_{y:|y|=2^{n_k}}B_y\right| \leq \left| \Sigma^{<g}\times 
\Sigma^{\leq p_j(n_k)} \right| = (2^g-1)(2^{p_j(n_k)+1}-1) < 
2^{p_j(n_k)+g+1} = 2^{2^{n_k}},
\]
which leads to a contradiction. Therefore, the lemma holds.
\end{proof}

Finally, the set $A$ is defined as the union $\bigcup_{i\in\nat}A_i$. 
By our construction, at each stage $k$, we need only $2^{O(n_k)}$ 
space to compute $A_k$ because $\MM_{U}$ uses $2^{O(n_k)}$ space to find 
the minimal string $y$ that satisfies Lemma \ref{claim5}. It thus 
follows that $A$ belongs to $\mathrm{ESPACE}$. 

Next, we want to show that $A$ is not in $\bqp_{tt}/\mathrm{poly\mbox{-}AUTO}$. 
Assume to the contrary that $A$ is in 
{\linebreak}
$\bqp_{tt}/\mathrm{poly\mbox{-}AUTO}$. 
There exists a polynomial-time truth-table query QTM $(M,N)$ that 
recognizes $A$ by polynomially-many nonadaptive queries with a 
polynomial advice function $h$. Lemma \ref{claim2} yields the existence of 
a polynomial $p$ and a constant $c$ satisfying that $\mathrm{C}^{p}(A[n]|h(n))
\leq 2^{n}-2^{n/3}+2n+ 2\log{n}+ c$ for any sufficiently large number $n$ in 
$\mathrm{Tower2}$. Choose two indices $l$ and $m$ such that $p(n)\leq p_l(n)$ 
and $|h(n)|\leq p_m(n)$ for all numbers $n\in\nat$. 
Now, take any numbers $i$ and $j$ such that $i\geq l$, $j\geq m$, $n_k\geq 2\log{n_k}+c$, and the number $k=\pair{i,j}$ is sufficiently large. By the choice of $n_k$, there exists a string 
$z\in\Sigma^{\leq p_m(n_k)}$ satisfying that $\mathrm{C}^{p_l}(A[n_k]|z) 
\leq 2^{n_k}-2^{n_k/3}+3n_k < 2^{n_k}-2^{n_k/4}$. This clearly contradicts 
Lemma \ref{claim5}. 
\end{proofof}

\paragraph{Acknowledgment.} 
The authors are grateful to Scott Aaronson for pointing out 
an early error and to Peter H{\o}yer for valuable comments 
at an early stage of this research. The first author also 
thanks to Sophie Laplante for a detailed presentation of her result.

\bibliographystyle{alpha}

\end{document}